\documentclass[12pt]{article}

\usepackage[english]{babel}
\usepackage{texdraw}
\begin{document}


\def\j{{\bf j}}
\def\tg{{ \rm tg}}



\begin{center}
{\Large \bf Noncommutative quantum analogs of constant curvature spaces
}
\end{center}

\begin{center}
{\bf N.A. Gromov} \\
 Department of Mathematics, \\
Komi Science Center UrD RAS \\
Kommunisticheskaya st., 24, Syktyvkar, 167982, Russia \\
E-mail: gromov@dm.komisc.ru
\end{center}

\begin{center}
{ \bf Abstract}
\end{center}

The  quantum    $N$-dimensional orthogonal vector  Cayley-Klein  spaces
 with different combinations of quantum structure and Cayley-Klein scheme of contractions and analytical continuations are described  for  multipliers, which include the first and the second powers  of contraction parameters in the transformation of deformation parameter.
The noncommutative  analogs of  constant curvature spaces are introduced.
The low dimensional spaces with $N=3,4$ are discussed in detail and all quantum analogs of the fibered spaces corresponding to nilpotent values of contraction parameters are given.
As a result the  wide variety of the quantum deformations  are obtained. 



\section{Introduction}

Spaces of constant curvature are among all spaces most symmetric ones.
Such $N$-dimensional   space  has maximal motion group depending on $N(N+1)/2$ group parameters.
Due to this reason  they have many fields of application both in mathematics and physics.
The uniform axiomatics of all  constant curvature spaces of arbitrary dimension was elaborated by
R.I.Pimenov \cite{P-65-3}, where the set of parameters taking real, imaginary and nilpotent values
was introduced.
Practically these spaces can be obtained from the spherical space by contractions and analytical continuations known as a Cayley-Klein scheme \cite{G-90}, where instead of zero tending contraction parameters of \cite{IW-53} are used nilpotent valued parameters.
Accordingly their motion groups are obtained from orthogonal group by Cayley-Klein contractions and analytical continuations.

Quantization theory of simple Lie groups  was suggested by L.D. Faddeev and his school \cite{FRT}.
Mathematically quantum groups  are noncommutative and noncocommutative deformations of Hopf algebras.
With the quantum orthogonal groups are connected quantum orthogonal vector spaces (or quantum Euclidean spaces), which are defined as an algebra functions with the set of generators satisfying certain commutation relations. Quantum orthogonal sphere of arbitrary dimension $S_q^{N}$ have been suggested in this paper \cite{FRT}.
The standard Podles quantum sphere \cite{P-1987} is connected with quantum unitary group and can be viewed as a quotient $SU_q(2)/U(1)$.
The example of a quantum 4-sphere motivated by Poisson structure was presented in \cite{BCT-2002}.
The twisted deformations which quantize the semiclassical structure defined by a generic element of the Cartan subalgebra results in twisted spheres \cite{AB-2002}.
The discrete family of Podles quantum spheres can be thought of as the family of q-deformed fuzzy spheres \cite{GMS-2001}.
So there are different approaches to a noncommutative spaces.

It is naturally to construct quantum analogs of Cayley-Klein groups and spaces with the  same contractions and analytical continuations scheme which holds in commutative case starting from the quantum orthogonal groups and quantum  vector spaces.
Two new aspects are added in noncommutative quantum case as compared with the commutative one.
First of all a transformation of the  deformation parameter $z=Jv$  under contraction need be  added for  quantum group \cite{CGST-1990} or space.
Secondly two mathematical structure (quantum deformation and Cayley-Klein scheme) need be combined.
This structures can be combined in a different ways \cite{G-1997}.
Different combinations of quantum structure
and Cayley-Klein scheme are described
with the help of permutations $\sigma. $

In the  papers \cite{GKK-97}--\cite{GK-06}, quantum orthogonal Cayley-Klein groups  in Cartesian  basis and the noncommutative  analogs of the possible commutative kinematics \cite{B-68}    
was constructed starting from the mathematical theory  of quantum groups and quantum vector spaces \cite{FRT}.
But analysis in the previous papers  was confined to the minimal multiplier $J$ in transformation of deformation parameter, which has the first power multiplication of contraction parameters.
This restriction imply 
that for certain combinations of quantum structure and Cayley-Klein scheme some contractions do
not exist.
In order to all Cayley-Klein contractions
 was possible for all permutations $\sigma $
it is necessary to regard non-minimal multiplier $J$, which include the first and the second powers  of contraction parameters.
In this paper we find non-minimal multipliers for the general
quantum  orthogonal  Cayley-Klein  spaces (Sec. 3) and describe in detail noncommutative quantum analogs of constant curvature spaces for $N=3$ (Sec. 4) and $N=4$ (Sec. 5).
In Sec. 2 we briefly remind the definition of the commutative  Cayley-Klein  spaces.

\section{Spaces of constant curvature }

The axiomatic description of the most symmetric spaces,
namely constant curvature spaces was given by R.I.Pimenov \cite{P-65-3}. All $3^N$ $N$-dimensional constant curvature spaces can be realized \cite{G-90}  on the spheres
\begin{displaymath}
S^N(j)=\{ \xi_1^2+j_1^2\xi_2^2+ \ldots +(1,N+1)^2\xi_{N+1}^2=1 \},
\end{displaymath}
in $(N+1)$-dimensional vector space $O^{N+1}(j)$,
where
\begin{displaymath}
(i,k)=\prod^{\max(i,k)-1}_{l=\min(i,k)}j_l, \quad
(k,k)\equiv 1,
\end{displaymath}
and each of parameters  $j_k$ takes three values $1,\iota_k,i, \  k=1, \ldots, N.$
Here $ {\iota}_k $ are nilpotent generators
 $  {\iota}_k^2=0,  $ with commutative law of multiplication
$ {\iota}_k{\iota}_m={\iota}_m{\iota}_k \not =0, \  k \neq m. $
We shall demand that the following heuristic rules be fulfilled:
for a real or complex $a$ the expression $a/\iota_k$ is defined only for $a=0$,
the expression $\iota_m/\iota_k$ is defined only for $m=k$, then $\iota_k/\iota_k=1$.

The intrinsic Beltrami coordinates
$r_k=\xi_{k+1}\xi_1^{-1},\; k=1,2,\ldots,N $
present the  coordinate system  for constant curvature space  $S^N(j)$, which coordinate lines
$r_k=const$ are geodesic.  $S^N(j)$ has positive curvature for $j_1=1,$
negative for $j_1=i$ and it is flat for  $j_1=\iota_1.$
For a flat space the Beltrami coordinates  coincide with the Cartesian ones.
Nilpotent value of the contraction parameter $j_k=\iota_k,\; k>1$ correspond to a fiber  space
with $(k-1)$-dimensional base $S_b=\left\{r_1,\ldots,r_{k-1}\right\}$ and $(N-k+1)$-dimensional  fiber $S_f=\left\{r_k,\ldots,r_{N}\right\}$.
In order to avoid terminological misunderstanding let us stress that we  have in view locally trivial fibering, which is defined by the projection $pr: S^N(\iota_k)  \rightarrow S_b $.
This fibering constitute foundation of the semi-Riemannian geometry
\cite{P-64-1}--\cite{P-68}, \cite{G-09}
and has nothing to do with the principal bundle.
Imaginary value of parameter $j_k=i$ correspond to pseudo-Riemannian space.
All nine constant curvature planes (or Cayley-Klein planes) are represented on Fig.1.

\vspace{6mm}

\begin{texdraw}
\drawdim{mm}
\move(0 0)
\lvec(130 0) \lvec(130 130) \lvec(0 130)
\lvec(0 0) \move (10 0) \lvec(10 130)
\move(50 0) \lvec(50 130) \move(90 0) \lvec(90 130)
\move(0 40) \lvec(130 40)
\move(0 80) \lvec(130 80)
\move(0 120) \lvec(130 120)
\textref h:C v:C \htext(5 100) {\footnotesize $1$}
\textref h:C v:C \htext(5 20) {\footnotesize $i$}
\textref h:C v:C \htext(5 60) {\footnotesize $\iota_2$}
\textref h:C v:C \htext(30 125) {\footnotesize $1$}
\textref h:C v:C \htext(70 125) {\footnotesize $\iota_1$}
\textref h:C v:C \htext(110 125) {\footnotesize $i$}
\textref h:C v:C \htext(2.3 122.5) {\footnotesize $j_2$}
\textref h:C v:C \htext(8 127.5) {\footnotesize $j_1$}
\move(0 130) \lvec(10 120)
\move(10 0)
\bsegment
\move(20 5)\arrowheadtype t:V \arrowheadsize l:2 w:1 \linewd 0.2 \ravec(0 30)
\move(5 20)\ravec(30 0)
\move(36 20) \larc r:25 sd:155 ed:205
\move(4 20) \larc r:25 sd:-25 ed:25
\move(20 52) \larc r:25 sd:245 ed:295
\move(20 -12) \larc r:25 sd:65 ed:115
\textref h:C v:C \htext(20 3) {\footnotesize anti de Sitter}
\textref h:C v:C \htext(23 35.5) {\footnotesize $r_1$}
\textref h:C v:C \htext(35 23) {\footnotesize $r_2$}
\move(20 20) \linewd 0.2 \lpatt(0.5 1) \rlvec(6 6)
\move(20 20) \rlvec(-6 -6) \move(20 20) \rlvec(6 -6) \move(20 20) \rlvec(-6 6)
\lpatt()
\esegment
\move(50 0)
\bsegment
\move(20 5)\arrowheadtype t:V \arrowheadsize l:2 w:1 \linewd 0.2 \ravec(0 30)
\move(5 20)\ravec(30 0)
\textref h:C v:C \htext(20 3) {\footnotesize Minkowski}
\textref h:C v:C \htext(23 35.5) {\footnotesize $r_1$}
\textref h:C v:C \htext(35 23) {\footnotesize $r_2$}
\move(12 8) \rlvec(0 24)
\move(28 8) \rlvec(0 24)
\move(8 12) \rlvec(24 0)
\move(8 28) \rlvec(24 0)
\move(20 20) \linewd 0.2 \lpatt(0.5 1) \rlvec(6 6)
\move(20 20) \rlvec(-6 -6) \move(20 20) \rlvec(6 -6) \move(20 20) \rlvec(-6 6)
\lpatt()
\esegment
\move(90 0)
\bsegment
\move(20 5)\arrowheadtype t:V \arrowheadsize l:2 w:1 \linewd 0.2 \ravec(0 30)
\move(5 20)\ravec(30 0)
\textref h:C v:C \htext(20 3) {\footnotesize de Sitter}
\textref h:C v:C \htext(23 35.5) {\footnotesize $r_1$}
\textref h:C v:C \htext(35 23) {\footnotesize $r_2$}
\move(52 20) \larc r:25 sd:155 ed:205
\move(-12 20) \larc r:25 sd:-25 ed:25
\move(20 38) \larc r:25 sd:245 ed:295
\move(20 2) \larc r:25 sd:65 ed:115
\move(20 20) \linewd 0.2 \lpatt(0.5 1) \rlvec(6 6)
\move(20 20) \rlvec(-6 -6) \move(20 20) \rlvec(6 -6) \move(20 20) \rlvec(-6 6)
\lpatt()
\esegment
\move(10 40)
\bsegment
\move(20 5)\arrowheadtype t:V \arrowheadsize l:2 w:1 \linewd 0.2 \ravec(0 30)
\move(5 20)\linewd 0.5 \ravec(30 0)\linewd 0.2
\textref h:C v:C \htext(20 3) {\footnotesize Newton(+)}
\textref h:C v:C \htext(23 35.5) {\footnotesize $r_1$}
\textref h:C v:C \htext(35 23) {\footnotesize $r_2$}
\move(36 20) \larc r:25 sd:155 ed:205
\move(4 20) \larc r:25 sd:-25 ed:25
\move(8 12)\linewd 0.5 \rlvec(24 0)
\move(8 28) \rlvec(24 0)
\esegment
\move(50 40)
\bsegment
\move(20 5)\arrowheadtype t:V \arrowheadsize l:2 w:1 \linewd 0.2 \ravec(0 30)
\move(5 20)\linewd 0.5 \ravec(30 0)\linewd 0.2
\textref h:C v:C \htext(20 3) {\footnotesize Galilei}
\textref h:C v:C \htext(23 35.5) {\footnotesize $r_1$}
\textref h:C v:C \htext(35 23) {\footnotesize $r_2$}
\move(8 12)\linewd 0.5 \rlvec(24 0)
\move(8 28) \rlvec(24 0)
\move(12 8)\linewd 0.2 \rlvec(0 24)
\move(28 8) \rlvec(0 24)
\esegment
\move(90 40)
\bsegment
\move(20 5)\arrowheadtype t:V \arrowheadsize l:2 w:1 \linewd 0.2 \ravec(0 30)
\move(5 20)\linewd 0.5 \ravec(30 0)\linewd 0.2
\textref h:C v:C \htext(20 3) {\footnotesize Newton(-)}
\textref h:C v:C \htext(23 35.5) {\footnotesize $r_1$}
\textref h:C v:C \htext(35 23) {\footnotesize $r_2$}
\move(52 20) \larc r:25 sd:155 ed:205
\move(-12 20) \larc r:25 sd:-25 ed:25
\move(8 12)\linewd 0.5 \rlvec(24 0)
\move(8 28) \rlvec(24 0)
\esegment
\move(10 80)
\bsegment
\move(20 5)\arrowheadtype t:V \arrowheadsize l:2 w:1 \linewd 0.2 \ravec(0 30)
\move(5 20)\ravec(30 0)
\move(36 20) \larc r:25 sd:155 ed:205
\move(4 20) \larc r:25 sd:-25 ed:25
\move(20 36) \larc r:25 sd:245 ed:295
\move(20 4) \larc r:25 sd:65 ed:115
\textref h:C v:C \htext(20 3) {\footnotesize Spherical}
\textref h:C v:C \htext(23 35.5) {\footnotesize $r_1$}
\textref h:C v:C \htext(35 23) {\footnotesize $r_2$}
\esegment
\move(50 80)
\bsegment
\move(20 5)\arrowheadtype t:V \arrowheadsize l:2 w:1 \linewd 0.2 \ravec(0 30)
\move(5 20)\ravec(30 0)
\textref h:C v:C \htext(20 3) {\footnotesize Euclid}
\textref h:C v:C \htext(23 35.5) {\footnotesize $r_1$}
\textref h:C v:C \htext(35 23) {\footnotesize $r_2$}
\move(8 12)\linewd 0.2 \rlvec(24 0)
\move(8 28) \rlvec(24 0)
\move(12 8)\linewd 0.2 \rlvec(0 24)
\move(28 8) \rlvec(0 24)
\esegment
\move(90 80)
\bsegment
\move(20 5)\arrowheadtype t:V \arrowheadsize l:2 w:1 \linewd 0.2 \ravec(0 30)
\move(5 20)\ravec(30 0)
\textref h:C v:C \htext(20 3) {\footnotesize Lobachevsky}
\textref h:C v:C \htext(23 35.5) {\footnotesize $r_1$}
\textref h:C v:C \htext(35 23) {\footnotesize $r_2$}
\move(52 20) \larc r:25 sd:155 ed:205
\move(-12 20) \larc r:25 sd:-25 ed:25
\move(20 52) \larc r:25 sd:245 ed:295
\move(20 -12) \larc r:25 sd:65 ed:115
\esegment
\end{texdraw}

\vspace{-6mm}

Fig.1. Nine constant curvature planes. Fibers are represented 
by thick lines.
Light cones in $(1+1)$ kinematics are drawn by dotted lines.

\vspace{6mm}

Part of constant curvature spaces $S^4(j)$ for
$ j_1=1,\iota_1,i, \; j_2=\iota_2,i,\; j_3=j_4=1 $ can be regarded  as $(1+3)$ space-time models or
 kinematics \cite{B-68},
if one interprets $r_1$ as the time axis  and the rest
as the space ones \cite{G-90},\cite{GK-06}.

\section{Quantum orthogonal groups and quantum  Cayley-Klein  spaces}

According to FRT theory \cite{FRT}, the algebra function on
quantum orthogonal group $ Fun(SO_q(N)) $ (or simply quantum orthogonal group $ SO_q(N) $) is the algebra of noncommutative polynomials of $ n^2 $
variables $ t_{ij}, i,j=1, \ldots, n, $ which are subject of commutation
relations
\begin{equation}
R_qT_1T_2=T_2T_1R_q
\label{3}
\end{equation}
and additional relations of $q$-orthogonality
\begin{equation}
TCT^t=C,\quad T^tC^{-1}T=C^{-1}.
\label{4}
\end{equation}
Here
 $ T_1=T \otimes I, \
 T_2=I \otimes T \in M_{n^2}({\bf C} \langle t_{ij} \rangle), $
$ T=(t_{ij})^n_{i,j=1} \in M_n({\bf C} \langle t_{ij} \rangle ), $
 $ I $ is unit matrix in  $ M_n({\bf C}), $
 $ C=C_0q^{\rho}, $  $ \rho=diag(\rho_1, \ldots, \rho_N), \
(C_0)_{ij}=\delta_{i'j}, \ i'=N+1-i, \  i,j=1, \ldots, N, $ that is
 $ (C)_{ij}=q^{\rho_{i'}}\delta_{i'j} $  and
$ C^{-1}=C, $
\begin{equation}
(\rho_1, \ldots, \rho_N)=
\left \{ \begin{array}{ccc}
     (n-\frac{1}{2}, n-\frac{3}{2}, \ldots , \frac{1}{2},0,-\frac{1}{2},
     \ldots , -n+\frac{1}{2}), \; N=2n+1 \\
     (n-1, n-2, \ldots, 1,0,0,-1, \ldots, -n+1), \; N=2n.
     \end{array} \right.
\label{5}
\end{equation}
The numerical matrix $ R_q $ is the well-known solution \cite{FRT}
of Yang-Baxter equation and its elements fulfils the role of  the structure constant of quantum group generators.

Let us remind the definition of the quantum vector space \cite{FRT}.

{\bf Definition 1.}
An algebra  $ O_q^N({\bf C}) $ with generators $ x_1,\ldots,x_N $ and
commutation relations
\begin{equation}
\hat R_q(x \otimes x)=qx \otimes x - \frac{q-q^{-1}}{1+q^{N-2}}
x^tCx  W_q,
\label{6}
\end{equation}
where
$ \hat R_q=PR_q, \; Pu \otimes v=v \otimes u,\;\forall u,v \in {\bf C}^n,\;
W_q=\sum^N_{i=1}q^{\rho_{i'}}e_i \otimes e_{i'},$
\begin{equation}
x^tCx=\sum^N_{i,j=1}x_iC_{ij}x_j=
\epsilon x_{n+1}^2+
\sum_{k=1}^n\left(
q^{-\rho_k}x_kx_{k'}+q^{\rho_k}x_{k'}x_k\right),
\label{7}
\end{equation}
$\epsilon =1 $ for $N=2n+1,$ $\epsilon =0 $ for $N=2n$
and vector $ (e_i)_k = \delta_{ik}, \enskip i,k=1,\ldots,N $
is called the algebra of functions on $N$-dimensional quantum
Euclidean space (or simply the quantum Euclidean space). 

The coaction of the quantum group $ SO_{q}(N) $ on the noncommutative
vector space $ O_q^N({\bf C}) $ is given by
\begin{displaymath}
\delta(x)=T \dot \otimes x, \quad
\delta(x_i)=\sum^n_{k=1}t_{ik} \otimes x_k, \;  i=1, \ldots,n
\end{displaymath}
and quadratic form  (\ref{7}) is invariant $x^tCx=\mbox{inv}$  with respect to
this coaction.

The matrix $C$ has non-zero elements only on the secondary diagonal. They are equal to unit in the commutative limit $q=1$. Therefore the quantum group
$ SO_q(N) $ and the quantum vector space $ O_q^N({\bf C}) $ are described
by equations (\ref{3})--(\ref{7}) in a skew-symmetric basis, where for $q=1$ the invariant form $x^tC_0x=\mbox{inv}$  is given by the matrix $C_0$ with the only non-zero elements  on the secondary diagonal which  are all equal to real units.

New generators $y=D^{-1}x$ of the quantum Euclidean  space $ O_q^N({\bf C}) $
in {\it  arbitrary} basis are obtained \cite{GKK-97},\cite{G-97}
with the help of non-degenerate matrix $D \in M_N$ and they are subject of
the commutation relations
\begin{displaymath}
\hat {R}(y \otimes y)=
q y \otimes y - \frac{\lambda}{1+q^{N-2}} y^tC'y W,
\end{displaymath}
where $ {\hat {R}}=(D \otimes D)^{-1}{\hat R_q}(D \otimes D),\;
W=(D \otimes D)^{-1}W_q,\; C'=D^tCD.$
The corresponding quantum group $ SO_q(N) $ is generated in arbitrary
basis by $ U=(u_{ij})^N_{i,j=1}, $ where  $ U=D^{-1}TD. $
The commutation relations of the new generators are
\begin{displaymath}
 \tilde{R}U_{1}U_{2} = U_{2}U_{1} \tilde{R}
\end{displaymath}
and $q$-orthogonality relations look as follows
\begin{displaymath}
U\tilde {C}U^t=\tilde {C}, \quad U^t(\tilde {C})^{-1}U=(\tilde {C})^{-1},
\end{displaymath}
where $ {\tilde {R}}=(D \otimes D)^{-1}{ R_q}(D \otimes D),\;
\tilde {C}=D^{-1}C(D^{-1})^t. $

In many  cases the most natural basis is the Cartesian basis, where the invariant form $\mbox{inv}=y^ty$ is given by the unit matrix $I.$
The transformation from the skew-symmetric  basis $x$
to the Cartesian basis $y$ is described by the matrix $D,$ which is a
solution of the following equation
\begin{equation}
D^tC_0D=I.
\label{12}
\end{equation}
This equation has many solutions. Take one of these, namely
\begin{equation}
D=\frac{1}{\sqrt{2}}
\left ( \begin{array}{ccc}
      I & 0 &  -i{\tilde C_0} \\
      0 & \sqrt{2} &  0 \\
      {\tilde C_0} & 0  &  iI
      \end{array} \right ),    \       N=2n+1,
\label{13}
\end{equation}
where $ {\tilde C_0} $ is the  $ n \times n $ matrix  with real units
on the secondary diagonal. For $N=2n$ the matrix $D$ is given by (\ref{13})
without the middle column and row.
The matrix (\ref{13}) provides one of the possible combinations of the quantum group structure and the Cayley-Klein  scheme of group contractions.
All other similar combinations  are given
by  the matrices $ D_{\sigma}=DV_{\sigma}, $  obtained from
(\ref{13}) by the right multiplication on the matrix $ V_{\sigma} \in M_{N} $ with elements
$ (V_{\sigma})_{ik}= \delta_{\sigma_{i},k}, $
where $ \sigma \in S(N) $ is a permutation of the $ N$-th order
\cite{GKK-2001}. The matrices $ D_{\sigma}$ are
solutions of  equation (\ref{12}).

We derive the quantum Cayley-Klein spaces with the same  transformation
of the Cartesian generators
$ y =\psi \xi, \; \psi ={\rm diag}(1,(1,2),\ldots,(1,N)) \in M_N, $
as in commutative case \cite{G-90}, \cite{G-97}.
The transformation  $z=Jv$ of the deformation parameter $q=e^z$ should be added in quantum case.

{\bf Definition 2.}
Algebra $ O^N_v(j;\sigma;{\bf C}) $ with Cartesian generators $ \xi_1,\ldots,\xi_N $ and
commutation relations
 \begin{equation}       
{{\hat R}}_{\sigma}(j) \xi \otimes \xi = e^{Jv} \xi \otimes \xi -
\frac{2\sinh Jv}{1+e^{Jv(N-2)}}\xi^tC_{\sigma}(j) \xi  {W}_{\sigma}(j),
\label{14}
\end{equation}    
where
\begin{displaymath}       
{\hat R}_{\sigma}(j)={\Psi}^{-1}{\hat R}_{\sigma}\Psi, \quad
W_{\sigma}(j)={\Psi}^{-1}{W}_{\sigma},
\end{displaymath}   
\begin{displaymath}
C_{\sigma}(j)=
{\psi} D^t_{\sigma} C D_{\sigma} \psi =
{\psi} V^{t}_{\sigma} D^t C D V_{\sigma} \psi,\;\;
\Psi = \psi \otimes \psi,
\end{displaymath}
is called the  $N$-dimensional quantum  Cayley-Klein vector space.

In explicit form  commutation relations (\ref{14}) are
\begin{displaymath}   
\xi_{\sigma_{k}}\xi_{\sigma_{m}}  =
\xi_{\sigma_{m}} \xi_{\sigma_{k}} \cosh Jv -
i\xi_{\sigma_{m}} \xi_{\sigma_{k'}}
\frac{(1,{\sigma_{k'}})}{(1,{\sigma_{k}})}\sinh Jv, \;
k<m<{k'}, \;  k \not = m',
 \end{displaymath}    
 \begin{equation}  
\xi_{\sigma_{k}}\xi_{\sigma_{m}}  =
\xi_{\sigma_{m}} \xi_{\sigma_{k}} \cosh Jv -
i\xi_{\sigma_{m'}} \xi_{\sigma_{k}}
\frac{(1,{\sigma_{m'}})}{(1,{\sigma_{m}})} \sinh Jv, \;
m'<k<m, \;  k \not = m',
\label{15-1}
 \end{equation}  
  \begin{displaymath}  
 [ \xi_{\sigma_{k}},\xi_{\sigma_{k'}}  ]
=  2i\epsilon
\sinh(\frac{Jv}{2}){(\cosh Jv)}^{n-k} \xi^2_{\sigma_{n+1}}
 \frac{(1,{\sigma_{n+1}})^{2}}{(1,{\sigma_{k}})(1,{\sigma_{k'}}) }  +
 \end{displaymath}  
\begin{equation}
+ i \frac{\sinh(Jv)}{(\cosh Jv)^{k+1}
(1,{\sigma_{k}})(1,{\sigma_{k'}}) }
    \sum^n_{m=k+1}
   {(\cosh   Jv)}^m  ((1,{\sigma_{m}})^{2}\xi^2_{\sigma_{m}} +
(1,{\sigma_{m'}})^{2} \xi^2_{\sigma_{m'}}),
\label{15}
 \end{equation}
where $k,m=1,2,\ldots,n,\; N=2n+1$ or $N=2n,\; k'=N+1-k $,
permutation
$\sigma=(\sigma_1,\ldots,\sigma_N)$
describes definite combination of the quantum group structure and Cayley-Klein scheme of group contraction.
The invariant form under the coaction of the corresponding quantum orthogonal group on the quantum Cayley-Klein space
$ O^N_v(j;\sigma;{\bf C}) $  is written as
\begin{displaymath}     
\mbox{inv}(j;\sigma) =
\Biggl(  \epsilon (1,\sigma_{n+1})^2\xi^2_{\sigma_{n+1}}
\frac{(\cosh Jv)^{n}}{\cosh(Jv/2)} +
\end{displaymath}          
\begin{equation}
+ \sum^n_{k=1} ( (1,\sigma_{k})^{2}\xi^2_{\sigma_{k}} +
(1,\sigma_{k'})^{2}\xi^2_{\sigma_{k'}}) {(\cosh Jv)}^{k-1}\Biggr)\cosh (Jv{ \rho_1}).
\label{16}
\end{equation}

{\bf Definition 3.}
 The quantum Euclidean space   $ O_q^N({\bf C}) $ with the antiinvolution  $x^{*}=C^tx$
 (or in components $x^{*}_k=q^{\rho_k}x_{k'}, \, k=1,\ldots,N$)
is called  the quantum real Euclidean space   $ O_q^N({\bf R}) $ \cite{FRT}.

Similar definition is hold for  quantum Cayley-Klein space.

{\bf Definition 4.}
The quantum  Cayley-Klein vector space $ O^N_v(j;\sigma;{\bf C}) $ with the antiinvolution
\begin{displaymath}
\xi^*_{\sigma_k}=\xi_{\sigma_k} \cosh Jv\rho_{k} +
i\xi_{\sigma_{k'}}\frac{(1,\sigma_{k'})}{(1,\sigma_{k})} \sinh Jv\rho_{k},\quad
\xi^*_{\sigma_{n+1}}=\xi_{\sigma_{n+1}},
\end{displaymath}
\begin{equation}
\xi^*_{\sigma_{k'}}=\xi_{\sigma_{k'}} \cosh Jv\rho_{k} -
i\xi_{\sigma_{k}}\frac{(1,\sigma_{k})}{(1,\sigma_{k'})} \sinh Jv\rho_{k}, \quad
k=1,\ldots,N
\label{inv}
\end{equation}
is called the  quantum real  Cayley-Klein vector space $ O^N_v(j;\sigma). $

The multiplier $J$ in the transformation  $z=Jv$ of the deformation parameter need be  chosen in such a way that all
indefinite relations in commutators, which appear under  nilpotent values of the contraction parameters are  canceled.

{\bf Theorem.}
{\it The quantum $N$-dimensional Cayley-Klein vector space
$ O^N_v(j;\sigma) $ exist for all possible contractions
$j_k=\iota_k, k=1,\ldots,N-1,$
if  multiplier $J$ in  transformation  of deformation parameter $z=Jv$ is taken in the form
\begin{equation}
J=J_0\bigcup J_1=J_0 \bigcup_{k} J_1^{(k)},
\label{16-4-1}
\end{equation}
where $J_0,J_1^{(k)},J_1$ are given by
(\ref{16-5}),(\ref{16-3}),(\ref{16-4}). }

{\bf Proof.}
As far as multipliers $(1,\sigma_k)$ and $(1,\sigma_{k'})$ enter symmetrically into the commutators (\ref{15-1}),(\ref{15}), we can put
$\sigma_k < \sigma_{k'} $ without  loss of  generality.
Then indefinite relations in commutators (\ref{15-1}) take the form $(1,\sigma_k)(1,\sigma_{k'})^{-1}=(\sigma_k,\sigma_{k'})$, where $k=1,2,\ldots,n $ for $N=2n+1$ and $k=1,2,\ldots,n-1 $ for $N=2n$.
They are eliminated by the multiplier
\begin{equation}
 J_0=\displaystyle{\bigcup_{k}(\sigma_k,\sigma_{k'})}.
\label{16-5}
\end{equation}
It has the first power multiplication of contraction parameters and is the {\it minimal} multiplier, which guarantees
the existence of the Hopf algebra structure for the associated quantum group $SO_v(N;j;\sigma).$ The analysis in the paper \cite{GK-06} was confined to this minimal case.

{\bf Definition 5.}
{ The ``union'' of two multipliers
  is understood as the 
multiplication of all parameters
$j_k,$ which occur  at least in one multiplier
   and the power of $j_k$ in the ``union'' is equal to its maximal power  in both multipliers,
for example, $(j_1j_2^2) \bigcup (j_2j_3)=j_1j_2^2j_3.$ }

If we take into account indefinite relations in commutators (\ref{15}), then we come to the non-minimal multiplier $J$, which consist of contraction parameters in the first and the second powers.
The indefinite relations in commutators (\ref{15}) have  the form
\begin{equation}
   \frac{\sum^n_{m=k+1} [(1,\sigma_{m})^{2} +
(1,\sigma_{m'})^{2}]}{(1,\sigma_k)(1,\sigma_{k'})}=
\frac{\sum^n_{m=k+1} (1,\sigma_{m})^{2}}{ (1,\sigma_k)^2(\sigma_k,\sigma_{k'})},
\label{16-1}
\end{equation}
for even $ N=2n$ and
\begin{equation}
   \frac{(1,\sigma_{n+1}) + \sum^n_{m=k+1} [(1,\sigma_{m})^{2} +
(1,\sigma_{m'})^{2}]}{(1,\sigma_k)(1,\sigma_{k'})}=
\frac{(1,\sigma_{n+1}) + \sum^n_{m=k+1} (1,\sigma_{m})^{2}}{ (1,\sigma_k)^2(\sigma_k,\sigma_{k'})},
\label{16-2}
\end{equation}
for odd $ N=2n+1$.
Let us introduce numbers
\begin{displaymath}   
i_k=\mbox{min}\{\sigma_{k+1},\ldots,\sigma_{n}\}, \;\; 
\end{displaymath}   
then $k$th expression in (\ref{16-1}) or (\ref{16-2}) is equal to
\begin{displaymath}          
   \frac{ (1,i_{k})^{2}}{ (1,\sigma_k)^2(\sigma_k,\sigma_{k'})}=\left\{
     \begin{array}{cc}
    (i_k,\sigma_k)^{-2}(\sigma_k,\sigma_{k'})^{-1},  & i_k < \sigma_k  \\
      (\sigma_k,i_{k})(i_k,\sigma_{k'})^{-1}, & \sigma_k  < i_k < \sigma_{k'}\\
    (\sigma_k,\sigma_{k'})(\sigma_{k'},i_k)^2,  & i_k > \sigma_{k'}
   \end{array}
   \right.
\end{displaymath}      
 and compensative multiplier  for this expression is as follows
\begin{equation}
 J_1^{(k)}=\left\{
     \begin{array}{cc}
(i_k,\sigma_k)^{2}(\sigma_k,\sigma_{k'}), & i_k < \sigma_k  \\
     (i_k,\sigma_{k'}), & \sigma_k  < i_k < \sigma_{k'}\\
  1,  & i_k > \sigma_{k'}.
   \end{array}
   \right.
    \label{16-3}
\end{equation}
For all expressions in (\ref{16-1}) or (\ref{16-2}) compensative multiplier $J_1$ is obtained by the union         
\begin{equation}
 J_1= \bigcup_{k} J_1^{(k)}
\label{16-4}
\end{equation}
Therefore the non-minimal  multiplier $J$ in the transformation  $z=Jv$ of the deformation parameter is given by (\ref{16-4-1}) and include the first and the second powers of contraction parameters.
$\triangle$

{\bf Definition 6.}
The quotient algebra $S_q^{N-1}$ of the algebra $O_q^N({\bf R})$ by the relation
$x^{*t}x=x^tCx=1$ is called the $(N-1)$-dimensional quantum orthogonal sphere \cite{FRT}.

We define quantum orthogonal Cayley-Klein sphere in a similar way.

{\bf Definition 7.}
The quotient algebra $S_v^{N-1}(j;\sigma) $  of the algebra $O_v^N(j;\sigma) $   by the relation
$ \mbox{inv}(j;\sigma)=1 $ (\ref{16})
 is called the $(N-1)$-dimensional quantum orthogonal Cayley-Klein sphere.

The quantum analogs of the intrinsic Beltrami coordinates on this quantum sphere
are given by  the sets of independent right or left generators
\begin{displaymath}
r_{\sigma_{i}-1}=\xi_{\sigma_i} \xi_1^{-1}, \quad
\hat{r}_{\sigma_{i}-1}=\xi_1^{-1} \xi_{\sigma_i} ,
\quad i=1,\ldots ,N, \quad i\neq k, \quad \sigma_k =1.
\end{displaymath}
The reason for  definition of right and left generators is the simplification of expressions for commutation relations.  It is possible to use only, say, right generators, but its commutators  are cumbersome, when all contraction parameters are not nilpotent.

\section{Quantum Cayley-Klein vector spaces $O^3_v({\bf j};\sigma)$ and
 orthogonal  spheres $S^2_v({\bf j};\sigma)$}

The 3-dimensional quantum Cayley-Klein vector spaces $O^3_v({\bf j};\sigma), {\bf j}=(j_1,j_2)$ are generated by
$\xi_{\sigma_l},\xi_{\sigma_2},\xi_{\sigma_3}$ with commutation relations
(see (\ref{15-1}),(\ref{15}))
\begin{displaymath} 
\xi_{\sigma_1}\xi_{\sigma_2} = \xi_{\sigma_2}\xi_{\sigma_1} \cosh(Jv)
-i\xi_{\sigma_2}\xi_{\sigma_3}\frac{(1,\sigma_3)}{(1,\sigma_1)} \sinh(Jv), \;
\end{displaymath} 
\begin{displaymath} 
\xi_{\sigma_2}\xi_{\sigma_3} = \xi_{\sigma_3}\xi_{\sigma_2} \cosh(Jv)
 - i\xi_{\sigma_1}\xi_{\sigma_2}\frac{(1,\sigma_1)}{(1,\sigma_3)} \sinh(Jv),
\end{displaymath} 
\begin{equation}
  [\xi_{\sigma_1},\xi_{\sigma_3}]   =
2i\xi_{\sigma_2}^2\frac{(1,\sigma_2)^2}{(1,\sigma_1)(1,\sigma_3)} \sinh(Jv/2)
 \label{3q-1}
\end{equation}
and has invariant form (\ref{16})
\begin{equation} 
\mbox{inv}(j;\sigma) =
\left((1,\sigma_1)^2\xi_{\sigma_1}^2 + (1,\sigma_3)^2\xi_{\sigma_3}^2\right)\cosh J\frac{v}{2} +
(1,\sigma_2)^2\xi_{\sigma_2}^2\cosh Jv.
\label{3-16}
\end{equation} 
The antiinvolution (\ref{inv}) of the Cartesian generators is written as
\begin{displaymath}
\xi^*_{\sigma_1}=\xi_{\sigma_1} \cosh Jv\rho_{1} +
i\xi_{\sigma_{3}}\frac{(1,\sigma_{3})}{(1,\sigma_{1})} \sinh Jv\rho_{1}, \quad
\xi^*_{\sigma_{2}}=\xi_{\sigma_{2}}
\end{displaymath}
\begin{equation}
\xi^*_{\sigma_{3}}=\xi_{\sigma_{3}} \cosh Jv\rho_{1} -
i\xi_{\sigma_{1}}\frac{(1,\sigma_{1})}{(1,\sigma_{3})} \sinh Jv\rho_{1}, \quad \rho_{1}=\frac{1}{2}, \; \rho_{2}=0.
\label{inv-3}
\end{equation}

By the  analysis of the multiplier (\ref{16-4-1}) for $N=3$
and commutation relations (\ref{3q-1}) of the quantum  space generators
we have find three permutations with  a
different multipliers  $J$,
 namely $J=j_1j_2$,
 for $  \sigma_0 =(1,2,3)$,
 $J=j_1$
for $ \sigma'=(1,3,2)$
and  $J=j_1^2j_2$
for $ \hat{\sigma} =(2,1,3). $

Orthogonal quantum 2-spheres
$
 S_v^{2}(j;\sigma)= O_v^{3}(j;\sigma)/\left\{\mbox{inv}(j;\sigma)=1\right\}
$
are characterized by the    right $r_k$ or left $\hat{r}_k$ quantum analogs of the intrinsic Beltrami coordinates, whose commutation relations can be obtained only for fixed permutation
$ \sigma $.
Let us discuss these cases of different permutations separately.

\subsection{Permutation $  \sigma_0 =(1,2,3)$, multiplier $J=j_1j_2$}

According to (\ref{3q-1}),(\ref{inv-3})
 the corresponding quantum Cayley-Klein vector space is characterized by the following commutation relations of generators and their involutions
 \begin{displaymath} 
 O^3_v(j;\sigma_0) =  \Biggl\{
\xi_{1}\xi_{2} = \xi_{2}\xi_{1} \cosh j_1j_2v
-i\xi_{2}\xi_{3}j_1j_2 \sinh j_1j_2v,
\end{displaymath} 
 \begin{displaymath} 
 \xi_{2}\xi_{3} = \xi_{3}\xi_{2} \cosh j_1j_2v
-i\xi_{1}\xi_{2}\frac{1}{j_1j_2} \sinh j_1j_2v, \quad
  [\xi_{1},\xi_{3}  ]  =
2i\xi_{2}^2 \frac{j_1^2}{j_1j_2} \sinh j_1j_2\frac{v}{2},
\end{displaymath} 
\begin{displaymath}
\xi_{1}^* = \xi_{1} \cosh j_1j_2\frac{v}{2}
+i\xi_{3}j_1j_2 \sinh j_1j_2\frac{v}{2},
\end{displaymath}
\begin{equation}
\xi_{3}^* = \xi_{3} \cosh j_1j_2\frac{v}{2}
-i\xi_{1}\frac{1}{j_1j_2} \sinh j_1j_2\frac{v}{2}, \quad
 \xi_{2}^* = \xi_{2}
\Biggr\}.
\label{1-inv-3}
\end{equation}
In the commutative case $(v=0)$ the nilpotent value of the first contraction parameter $j_1=\iota_1$ and $j_2=1$
gives the semi-Euclidean space with one dimensional base $\{\xi_1\}$ and two dimensional fiber $\{\xi_2,\xi_3\}$.
The noncommutative deformations of this fiber semi-Euclidean space is obtained by putting $j_1=\iota_1$ in  (\ref{1-inv-3}), namely
\begin{equation}
O^3_v(\iota_1;\sigma_0) =
\Biggl\{
 [\xi_{1},\xi_{2}] = [\xi_{1},\xi_{3}]=0, \; 
[\xi_{3},\xi_{2}] = iv\xi_{1}\xi_{2},\;
\xi_1^*=\xi_1,\; \xi_2^*=\xi_2,\; \xi_3^*=\xi_3-i\xi_1\frac{v}{2}
\Biggr\}.
\label{31}
\end{equation}

For $v=0$ contraction $j_1=1,j_2=\iota_2$ transforms Euclidean space $E_3$ to the space with two dimensional base  $\{\xi_1,\xi_2\}$ and one dimensional fiber $\{\xi_3\}$. Its quantum analogs is obtained by this contraction of commutators (\ref{1-inv-3})
\begin{equation}
O^3_v(\iota_2;\sigma_0) =
\Biggl\{
 [\xi_{1},\xi_{2}] =0, \; [\xi_{3},\xi_{1}] = iv\xi_{2}^2, \;
[\xi_{3},\xi_{2}] = iv\xi_{1}\xi_{2} 
\Biggr\}.
\label{32}
\end{equation}
The involutive generators $\xi_k^*,\; k=1,2,3 $ are the same as in (\ref{31}).

The invariant form for  permutation $\sigma_0$ is obtained from (\ref{3-16})
\begin{equation}
\mbox{inv}(j;\sigma_0) =
(\xi_1^2 + j_1^2j_2^2\xi_3^2)\cosh j_1j_2\frac{v}{2} + j_1^2\xi_2^2\cosh j_1j_2v.
\label{30}
\end{equation}
Orthogonal quantum 2-sphere
$
 S_v^{2}(j;\sigma_0)= O_v^{3}(j;\sigma_0)/\left\{\mbox{inv}(j;\sigma_0)=1\right\}
$
is described  by the  quantum analogs of  Beltrami coordinates with commutation relations
\begin{displaymath} 
 S_v^{2}(j;\sigma_0)= \Biggl\{
 r_1=\hat{r}_1(\cosh j_1j_2v -ir_2j_1j_2\sinh j_1j_2v), \;
 \end{displaymath} 
\begin{equation} 
r_2-\hat{r}_2=2i\hat{r}_1 r_1\sinh j_1j_2\frac{v}{2},\;\;
\hat{r}_1r_2=(\hat{r}_2\cosh j_1j_2v -i\frac{1}{j_1j_2}\sinh j_1j_2v)r_1.
\Biggl\}
 \label{1-3q-6}
\end{equation} 
For $j_1=\iota_1, j_2=1$ we obtain from (\ref{1-3q-6}) that the left   generators are equal to  the right $\hat{r}_1=r_1, \hat{r}_2=r_2 $ and
 the orthogonal quantum plane has the following commutation relations
\begin{equation}
 S_v^{2}(\iota_1;\sigma_0)= \biggl\{ [r_2,r_1]=ivr_1 \biggl\}. 
\label{3q-7}
\end{equation}

For $j_2=\iota_2, j_1=1$ we obtain from (\ref{1-3q-6}) the quantum analog of the
cylinder with $r_1=\hat{r}_1$ being its cyclic generatrix and $ \hat{r}_2=r_2-ivj_1^2r_1^2$.
If $j_1=i$, then cylinder has hyperbolic generatrix.
 The Beltrami generators of noncommutative  cylinder has the following commutation relations
  \begin{equation}
 S_v^{2}(\iota_2;\sigma_0)= \biggl\{ [r_2,r_1]=ivr_1(1+j_1^2r_1^2) \biggl\}.
\label{3q-8}
\end{equation}
For $j_1=\iota_1, j_2=\iota_2 $ the quantum Galilei 2-plane is given by (\ref{3q-7}).

\subsection{Permutation $  \sigma' =(1,3,2)$, multiplier $J=j_1$}

As it follows from (\ref{3q-1}),(\ref{inv-3}) for permutation $\sigma'$
the  commutation relations of generators and their involutions
 of the  quantum Cayley-Klein vector space $ O^3_v(j;\sigma')$ is described by
\begin{displaymath} 
 O^3_v(j;\sigma') =  \Biggl\{
\xi_{1}\xi_{3} = \xi_{3}\xi_{1} \cosh j_1v
-i\xi_{3}\xi_{2}j_1 \sinh j_1v,
\end{displaymath} 
\begin{displaymath} 
 \xi_{3}\xi_{2} = \xi_{2}\xi_{3} \cosh j_1v
-i\xi_{1}\xi_{3}\frac{1}{j_1} \sinh j_1v,\quad
  [\xi_{1},\xi_{2}    ]  =
2i\xi_{3}^2 \frac{j_1^2j_2^2}{j_1} \sinh j_1\frac{v}{2},
 \end{displaymath} 
\begin{displaymath}
\xi_{1}^* = \xi_{1} \cosh (j_1\frac{v}{2}
+i\xi_{3}j_1 \sinh j_1\frac{v}{2},
\end{displaymath}
\begin{equation}
\xi_{2}^* = \xi_{2} \cosh j_1\frac{v}{2}
-i\xi_{1}\frac{1}{j_1} \sinh j_1\frac{v}{2}, \quad
 \xi_{3}^* = \xi_{3}
\Biggr\},
\label{1'}
\end{equation}
%
%
For $j_1=\iota_1$ the quantum semi-Euclidean space $O^3_v(\iota_1;\sigma') $ is connected with the space $O^3_v(\iota_1;\sigma_0)$  (\ref{31}) by the replacement
$\xi_2 \rightarrow \xi_3$ and vice-versa, i.e. by the renumbering  of the fiber generators. So it can not be regarded as the independent nonequivalent deformation of the fiber space.
For  $j_1=1,j_2=\iota_2$ in (\ref{1'}) we obtain the noncommutative deformation of the fiber space with 2-dimensional commutative base  $\{\xi_1,\xi_2\}$ and 1-dimensional fiber $\{\xi_3\}$
\begin{displaymath} 
O^3_v(\iota_2;\sigma') =
\Biggl\{
 \; \xi_{1}\xi_{3} = \xi_{3}\xi_{1}\cosh v - i\xi_{3}\xi_{2}\sinh v, \;
 \xi_{3}\xi_{2} = \xi_{2}\xi_{3}\cosh v - i\xi_{1}\xi_{3}\sinh v,
 \end{displaymath}
\begin{equation}
 [\xi_{1},\xi_{2}] =0,\; \xi_3^*=\xi_3,\;
 \xi_1^*=\xi_1 \cosh \frac{v}{2} + i\xi_2\sinh \frac{v}{2},\;
  \xi_2^*=\xi_2 \cosh \frac{v}{2} - i\xi_1\sinh \frac{v}{2}
  \Biggr\}.
  \label{2'}
\end{equation}

The invariant forms for  permutation $\sigma'$ is obtained from  (\ref{3-16})
\begin{equation}
\mbox{inv}(j;\sigma') =
(\xi_1^2 + j_1^2\xi_2^2)\cosh j_1\frac{v}{2} + j_1^2j_2^2\xi_3^2\cosh j_1v.
\label{3'}
\end{equation}
Orthogonal quantum 2-sphere
$
 S_v^{2}(j;\sigma')= O_v^{3}(j;\sigma')/\left\{\mbox{inv}(j;\sigma')=1\right\}
$
has two  Beltrami generators with commutation relations
\begin{displaymath} 
 S_v^{2}(j;\sigma')= \Biggl\{
 \hat{r}_2=(\cosh j_1v +i\hat{r}_1j_1\sinh j_1v)r_2, \;
\end{displaymath} 
\begin{equation}
r_1-\hat{r}_1=2i\hat{r}_2 r_2\frac{j_1^2j_2^2}{j_1}\sinh j_1\frac{v}{2},\;\;
\hat{r}_2r_1=(\hat{r}_1\cosh j_1v -i\frac{1}{j_1}\sinh j_1v)r_2,
\Biggl\}.
\label{4'}
\end{equation}

For $j_1=\iota_1$ in (\ref{4'}) the quantum plane $S^2_v(\iota_1;\sigma') $ is connected with the quantum plane $S^2_v(\iota_1;\sigma_0)$  (\ref{3q-7}) by the replacement
$r_1 \rightarrow r_2$ and vice-versa,  so it can not be regarded as the  nonequivalent deformation of the orthogonal quantum plane.

For $j_2=\iota_2, j_1=1$ we obtain from (\ref{4'})
the quantum analog of the cylinder with
$r_1=\hat{r}_1$ being its cyclic ($j_1=1$) or hyperbolic ($j_1=i$) generatrix and
$ \hat{r}_2=(\cosh j_1v +i r_1j_1\sinh j_1v)r_2 $.
This quantum cylinder is described by
\begin{equation}
 S_v^{2}(\iota_2;\sigma')= \biggl\{
[r_1,r_2]=i(r_2+j_1^2r_1r_2r_1)\frac{1}{j_1}\tanh j_1v
 \biggl\}
\label{5'}
\end{equation}
and can be regarded as noncommutative deformation of the semi-Riemannian spaces  with 1-dimensional base $\{r_1\}$ and 1-dimensional fiber $\{r_2\}$. 
Spaces (\ref{2'}) and (\ref{5'}) give an example of contraction, when deformation parameter remain unchanged.   
Physically these spaces can be interpreted as quantum analogs of the $(1+1)$ nonrelativistic  Newton kinematics with constant curvature.

\subsection{Permutation $ \hat{\sigma} =(2,1,3))$, multiplier $J=j_1^2j_2$}

The  commutation relations of generators and their involutions
 of the  quantum Cayley-Klein vector space $ O^3_v(j;\hat{\sigma}) $
follow from (\ref{3q-1}),(\ref{inv-3})
\begin{displaymath} 
 O^3_v(j;\hat{\sigma}) =  \Biggl\{
\xi_{2}\xi_{1} = \xi_{1}\xi_{2} \cosh(j_1^2j_2v)
-i\xi_{1}\xi_{3}j_2 \sinh(j_1^2j_2v),
 \end{displaymath} 
   \begin{displaymath} 
 \xi_{1}\xi_{3} = \xi_{3}\xi_{1} \cosh(j_1^2j_2v)
-i\xi_{2}\xi_{1}\frac{1}{j_2} \sinh(j_1^2j_2v),\quad
  [\xi_{2},\xi_{3}]  =
2i\xi_{1}^2 \frac{1}{j_1^2j_2} \sinh(j_1^2j_2v/2),
\end{displaymath}
\begin{displaymath}
\xi_{2}^* = \xi_{2} \cosh(j_1^2j_2v/2)
+i\xi_{3}j_2 \sinh(j_1^2j_2v/2),
\end{displaymath}
\begin{equation}
\xi_{3}^* = \xi_{3} \cosh(j_1^2j_2v/2)
-i\xi_{2}\frac{1}{j_2} \sinh(j_1^2j_2v/2), \quad
 \xi_{1}^* = \xi_{1}
\Biggr\}.
 \label{3q-2}
\end{equation}

The nilpotent value of the first contraction parameter $j_1=\iota_1$ and $j_2=1$ in (\ref{3q-2})
gives the new quantum semi-Euclidean space
\begin{equation}
O^3_v(\iota_1;\hat{\sigma}) =
\Biggl\{
[\xi_{1},\xi_{2}] = [\xi_{1},\xi_{3}]=0,\;
[\xi_{2},\xi_{3}] = iv\xi_{1}^2,\;
\xi^*_k=\xi_k,\; k=1,2,3
\Biggr\},
 \label{3q-3}
\end{equation}
which is not isomorphic to (\ref{31}).

For $j_1=1,j_2=\iota_2$ quantum space $O^3_v(\iota_2;\hat{\sigma})$ is transformed to $O^3_v(\iota_2;\sigma_0)$
by the replacement of the base generators
$\xi_1 \rightarrow \xi_2$ and vice-versa, therefore  it does not present new noncommutative deformation.

For $j_1=\iota_1,j_2=\iota_2$ commutation relations of generators are given by (\ref{3q-3}).

The invariant form for  permutation $\hat{\sigma} $ is given by (\ref{3-16})
\begin{equation}
 \mbox{inv}(j;\hat{\sigma}) =
j_1^2(\xi_2^2 + j_2^2\xi_3^2)\cosh j_1^2j_2\frac{v}{2} + \xi_1^2\cosh j_1^2j_2v.
\label{1-3q}
\end{equation}
Orthogonal quantum 2-sphere
$
 S_v^{2}(j;\hat{\sigma})= O_v^{3}(j;\hat{\sigma})/\left\{\mbox{inv}(j;\hat{\sigma})=1\right\}
$
is characterized by the  commutation relations
\begin{displaymath} 
 S_v^{2}(j;\hat{\sigma})= \Biggl\{
 \hat{r}_1=r_1\cosh j_1^2j_2v -ir_2j_2\sinh j_1^2j_2v, \;
\end{displaymath} 
\begin{equation}
r_2=\hat{r}_2\cosh j_1^2j_2v -i\hat{r}_1\frac{1}{j_2}\sinh j_1^2j_2v,\;\;
\hat{r}_1r_2-\hat{r}_2r_1=2i\frac{1}{j_1^2j_2}\sinh j_1^2j_2\frac{v}{2}
\Biggl\}.
 \label{3q-6}
\end{equation}

For $j_1=\iota_1, j_2=1$ we obtain from (\ref{3q-6}) that the left   generators are equal to  the right $\hat{r}_1=r_1, \hat{r}_2=r_2 $ and
 orthogonal quantum plane is as follows
\begin{equation}
S_v^{2}(\iota_1;\hat{\sigma})= \biggl\{ [r_1,r_2]=iv \biggl\}.
\label{1-3q-7}
\end{equation}
This is the simplest deformation of the Euclid plane because commutator is proportional to the number $iv$, instead of operator in (\ref{3q-7}).

For $j_2=\iota_2, j_1=1$ the quantum cylinder $S_v^{2}(\iota_2;\hat{\sigma})$ is given by
\begin{equation}
S_v^{2}(\iota_2;\hat{\sigma})= \biggl\{ [r_1,r_2]=iv(1+j_1^2r_1^2) \biggl\}.
\label{2-3q-7}
\end{equation}
For $j_1=\iota_1, j_2=\iota_2 $ the simplest quantum deformation of the Galilei plane is given by
(\ref{1-3q-7}).
%
%
%

\section{Quantum  spaces $O^4_v({\bf j};\sigma)$ and $S^3_v({\bf j};\sigma)$}

Quantum vector spaces $O^4_v({\bf j};\sigma), {\bf j}=(j_1,j_2,j_3)$ are generated by
$\xi_{\sigma_l},\; l=1,\ldots,4$ with commutation relations $(k=2,3)$
\begin{displaymath} 
O^4_v(\j;\sigma)=\Biggl\{
\xi_{\sigma_1}\xi_{\sigma_k} = \xi_{\sigma_k}\xi_{\sigma_1} \cosh(Jv)
-i\xi_{\sigma_k}\xi_{\sigma_{1'}}\frac{(1,\sigma_{1'})}{(1,\sigma_1)} \sinh(Jv), \;
\end{displaymath} 
\begin{displaymath} 
\xi_{\sigma_k}\xi_{\sigma_{1'}} = \xi_{\sigma_{1'}}\xi_{\sigma_k} \cosh(Jv)
 - i\xi_{\sigma_1}\xi_{\sigma_k} \frac{(1,\sigma_1)}{(1,\sigma_{1'})} \sinh(Jv),\quad
 [\xi_{\sigma_2},\xi_{\sigma_{2'}}]  =0,
\end{displaymath} 
\begin{equation}
 [ \xi_{\sigma_1},\xi_{\sigma_{1'}} ] =
i\Biggl( \xi_{\sigma_2}^2 (1,\sigma_2)^2
+\xi_{\sigma_{2'}}^2 (1,\sigma_{2'})^2 \Biggr) \frac{\sinh(Jv)}{(1,\sigma_1)(1,\sigma_{1'})} \Biggr\},
\label{4q-1}
\end{equation}
where $\sigma_{1'}=\sigma_4, \sigma_{2'}=\sigma_3 $.
The antiinvolution (\ref{inv}) of the Cartesian generators is written as
\begin{displaymath}
\xi^*_{\sigma_1}=\xi_{\sigma_1} \cosh Jv +
i\xi_{\sigma_{4}}\frac{(1,\sigma_{4})}{(1,\sigma_{1})} \sinh Jv, \quad
\xi^*_{\sigma_{2}}=\xi_{\sigma_{2}},
\end{displaymath}
\begin{equation}
\xi^*_{\sigma_{4}}=\xi_{\sigma_{4}} \cosh Jv -
i\xi_{\sigma_{1}}\frac{(1,\sigma_{1})}{(1,\sigma_{4})} \sinh Jv, \quad
\xi^*_{\sigma_{3}}=\xi_{\sigma_{3}},
\label{6-2}
\end{equation}
since according with (\ref{5}) $\rho_1=1, \rho_2=0$.

By the  analysis of the multiplier (\ref{16-4-1}) for $N=4$
and commutation relations (\ref{4q-1}) of the quantum  space generators
we have find   minimal multiplier $J=(\sigma_1,\sigma_{1'})$, which takes three values
$J_0=(1,1')=j_1j_2j_3$ for permutation $\sigma_0 =(1,2,3,4)$,
 $J_{I}=(1,2')=j_1j_2$ for $\sigma_{I} =(1,2,4,3)$,
 $J_{II}=(1,2)=j_1$ for $\sigma_{II} =(1,3,4,2)$, i.e. for permutations with $\sigma_1=1$
and three non-minimal multipliers $J=(1,\sigma_1)(1,\sigma_{1'})$, namely
$J_{III}=(1,2')(1,1')=j_1^2j_2^2j_3$ for $\sigma_{III} =(3,1,2,4)$,
$J_{IV}=(1,2)(1,1')=j_1^2j_2j_3$ for $\sigma_{IV} =(2,1,3,4)$,
 $J_{V}=(1,2)(1,2')=j_1^2j_2$ for $\sigma_{V} =(2,1,4,3)$, i.e. for  permutations with $\sigma_1\neq 1$.

\subsection{Quantum fibered spaces $O^4_v({\bf j};\sigma)$ }

We shall not describe all six  combinations of Cayley-Klein and quantum structures in  full details but shall concentrate our attention on fibered spaces, which correspond to nilpotent values of the  contraction parameters.
The careful analysis of the commutation relations (\ref{4q-1}) for the above mentioned permutations
and nilpotent value of the first contraction parameter $j_1=\iota_1, j_2=j_3=1$ gives two nonisomorphic
quantum fibered spaces with 1-dimensional base $\{\xi_1\}$ and 3-dimensional fiber $\{\xi_2,\xi_3,\xi_4 \}$.
These  quantum fibered spaces are obtained for permutations $\sigma_0, \sigma_{III}$ and are characterized
by the following nonzero commutation relations
\begin{equation}
O^4_v(\iota_1;\sigma_0) =
\Biggl\{
  [\xi_{4},\xi_{k}] =iv\xi_{1}\xi_{k}.\; k=2,3
\Biggr\}, \quad
O^4_v(\iota_1;\sigma_{III}) =
\Biggl\{
  [\xi_{3},\xi_{4}] =iv\xi_{1}^2
\Biggr\}.
\label{bn-0}
 \end{equation}
In both cases base generator  commute with all fiber generators and the last ones are not closed under
commutation relations. The same properties are hold for $j_1=\iota_1, j_2=\iota_2, j_3=1$, $j_1=\iota_1, j_3=\iota_3, j_2=1$,
$j_1=\iota_1, j_2=\iota_2, j_3=\iota_3$, i.e. for successive enclosed   projections or  repeatedly  fibered spaces.

When the second contraction parameter takes nilpotent value $ j_2=\iota_2, j_1=j_3=1$, then the fibered commutative space with 2-dimensional base $\{\xi_1,\xi_2\}$ and 2-dimensional fiber $\{\xi_3,\xi_4\}$ is obtained.  There are three its nonisomorphic
noncommutative analogs, which are given by (\ref{4q-1}) for permutations $\sigma_0, \sigma_{II}, \sigma_{III}$. Their nonzero commutators are as follows
\begin{displaymath} 
O^4_v(\iota_2;\sigma_0) =
\Biggl\{
  [\xi_{4},\xi_{k}] =iv\xi_{1}\xi_{k},\; k=2,3
\Biggr\}, \quad
O^4_v(\iota_2;\sigma_{III}) =
\Biggl\{
  [\xi_{3},\xi_{4}] =iv\xi_{1}^2
\Biggr\},
\end{displaymath} 
\begin{displaymath} 
O^4_v(\iota_2;\sigma_{II}) =
\Biggl\{
 \xi_{1}\xi_{k} = \xi_{k}(\xi_{1}\cosh v - i\xi_{2}\sinh v),\;
\end{displaymath} 
 \begin{equation}
  \xi_{k}\xi_{2} =(\xi_{2}\cosh v - i\xi_{1}\sinh v) \xi_{k}, \; k=3,4
  \Biggr\}.
 \label{bn-1}
 \end{equation}
The base generators commute for all permutations.
The fiber generators commute only for  permutation $\sigma_{II}$.
The base generators do not commute with the fiber generators for all permutations.
The fiber generators are not closed with respect commutation relations for $\sigma_{0}$ and $\sigma_{III}$.
The same properties are hold for $j_1=1,  j_2=\iota_2, j_3=\iota_3 $.

The fibered commutative space with 3-dimensional base $\{\xi_1,\xi_2\,\xi_3\}$ and 1-dimensional fiber $\{\xi_4\}$ is obtained for nilpotent value of the third parameter $ j_3=\iota_3, j_1=j_2=1 $.
We have find two nonisomorphic quantum fibered spaces for this values of contraction parameters, which are given by (\ref{4q-1}) for permutations $\sigma_0, \sigma_{II}$ and are characterized by the nonzero commutation relations
\begin{displaymath} 
O^4_v(\iota_3;\sigma_0) =
\Biggl\{
  [\xi_{4},\xi_{k}] =iv\xi_{1}\xi_{k},\; k=2,3,\;
  [\xi_{1},\xi_{4}] =iv(\xi_{2}^2 + \xi_{3}^2)
\Biggr\},
\end{displaymath} 
\begin{displaymath} 
O^4_v(\iota_3;\sigma_{II}) =
\Biggl\{
 \xi_{1}\xi_{k} = \xi_{k}(\xi_{1}\cosh v - i\xi_{2}\sinh v),\;
\end{displaymath} 
 \begin{equation}
 \xi_{k}\xi_{2} =(\xi_{2}\cosh v - i\xi_{1}\sinh v) \xi_{k}, \; k=3,4,\;\;
  [\xi_{1},\xi_{2}] =i\xi_{3}^2\sinh v
  \Biggr\}.
  \label{bn-2}
 \end{equation}
The base generators commute for  permutation $\sigma_0$, but do not commute for  permutation $\sigma_{II}$.
In the last case they are closed with respect commutation relations.

In general quantum  spaces $O^4_v(j;\sigma)$ have commutative base for all permutations, when fibering is defined by
$j_1=\iota_1$ or $j_2=\iota_2$. When fibering is defined by $j_3=\iota_3$ the 3-dimensional base is also commutative for
permutation $\sigma_0$.
The only exception is the quantum space $O^4_v(\iota_3;\sigma_{II})$, where  3-dimensional base is noncommutative, but closed with respect commutation relations.
For all permutations and all nilpotent values of contraction parameters the fibers are noncommutative and nonclosed except for
$O^4_v(\iota_2;\sigma_{II})$, where both 2-base and 2-fiber are commutative.

The antiinvolution of generators is easily obtained from general expressions (\ref{6-2}).
For $O^4_v(\iota_1;\sigma_0)$, $O^4_v(\iota_2;\sigma_0)$, $O^4_v(\iota_3;\sigma_0)$ we have
$\xi^*_m=\xi_m,\;  m=1,2,3,\; \xi^*_4=\xi_m-iv\xi_1$.
For $O^4_v(\iota_1;\sigma_{III}), O^4_v(\iota_2;\sigma_{III})$ antiinvolution looks  very simple
$\xi^*_k=\xi_k, k=1,2,3,4$.
The most complicate antiinvolution
\begin{displaymath}
\xi^*_{1}=\xi_{1} \cosh j_1v +
i\xi_{2}j_1 \sinh j_1v, \quad
\end{displaymath}
\begin{displaymath}
\xi^*_{2}=\xi_{2} \cosh j_1v -
i\xi_{1}\frac{1}{j_1} \sinh j_1v, \quad
\xi^*_{s}=\xi_{s},\; s=3,4.
\end{displaymath}
has quantum spaces $O^4_v(\iota_2;\sigma_{II}), O^4_v(\iota_3;\sigma_{II})$.

\subsection{Quantum deformations of constant curvature spaces  $S^3_v({\bf j};\sigma)$}

The invariant form of $O^4_v(j;\sigma)$ is given by (\ref{16}) for $N=4$
\begin{displaymath} 
\mbox{inv}(j;\sigma) = \left[
(1,\sigma_1)^2\xi_{\sigma_1}^2 + (1,\sigma_4)^2\xi_{\sigma_4}^2 \right.
\left. + \left((1,\sigma_2)^2\xi_{\sigma_2}^2 + (1,\sigma_3)^2\xi_{\sigma_3}^2 \right)
\cosh Jv
\right]\cosh Jv.
\end{displaymath}
The 3-dimensional quantum orthogonal sphere $S_v^{3}(j;\sigma) $ is obtained
as the quotient of $O_v^4(j;\sigma) $ by $ \mbox{inv}(j;\sigma)=1 $.
It is described by the noncommutative sets of  right and left space generators
$r_k=\xi_{k+1}\xi^{-1},\; \hat{r}_k=\xi^{-1}\xi_{k+1},\; k=1,2,3$. For different permutations $\sigma_0, \sigma_I,\ldots,\sigma_V $
these spheres are
\begin{equation}
 S_v^{3}(j;\sigma_0)= \Biggl\{
 r_1r_2=r_2r_1,\;
 \hat{r}_mr_3=\left(\hat{r}_3\cosh J_0v -i\frac{1}{J_0}\sinh J_0v\right)r_m,\; m=1,2 \Biggl\},
\label{4q-45}
\end{equation} 
 where
\begin{displaymath} 
\hat{r}_m= \left(\cosh J_0v + i\hat{r}_3 J_0 \sinh J_0v \right) r_m,\; m=1,2,\;
\end{displaymath} 
\begin{displaymath} 
r_3-\hat{r}_3=ij_1^2\left(\hat{r}_1 r_1 +j_2^2\hat{r}_2 r_2 \right)
 \frac{1}{J_0}\sinh J_0v,\;\; J_0=j_1j_2j_3.
   \end{displaymath} 
\begin{equation}   
 S_v^{3}(j;\sigma_I)= \Biggl\{
 r_1r_3=r_3r_1,\;
 \hat{r}_mr_2=\left(\hat{r}_2\cosh J_Iv -i\frac{1}{J_I}\sinh J_Iv\right)r_m,\; m=1,3 \Biggl\},
  \label{4q-25}
\end{equation} 
 where
\begin{displaymath} 
\hat{r}_m=\left( \cosh J_Iv + i\hat{r}_2 J_I \sinh J_Iv \right)r_m,\; m=1,3,\;
\end{displaymath} 
\begin{displaymath} 
r_2-\hat{r}_2=ij_1^2\left(\hat{r}_1 r_1 +j_2^2j_3^2\hat{r}_3 r_3 \right)
 \frac{1}{J_I}\sinh J_Iv,\;\; J_I=j_1j_2.
\end{displaymath} 
\begin{equation}   
 S_v^{3}(j;\sigma_{II})= \Biggl\{
 r_2r_3=r_3r_2,\;
 \hat{r}_mr_1=\left(\hat{r}_1\cosh J_{II}v -i\frac{1}{J_{II}}\sinh J_{II}v \right)r_m,\; m=2,3 \Biggl\},
 \label{4q-16}
\end{equation} 
 where
\begin{displaymath} 
\hat{r}_m=\left( \cosh J_{II}v + i\hat{r}_1 J_{II} \sinh J_{II}v \right)r_m,\; m=2,3,\;
\end{displaymath} 
\begin{displaymath} 
r_1-\hat{r}_1=ij_1^2j_2^2\left(\hat{r}_2 r_2 +j_3^2\hat{r}_3 r_3 \right)
 \frac{1}{J_{II}}\sinh J_{II}v,\;\; J_{II}=j_1.
\end{displaymath} 
Commutation relations for nonminimal multipliers are more simple
\begin{displaymath} 
 S_v^{3}(j;\sigma_{III})= \Biggl\{
 [r_1,r_2]=[r_1,r_3]=0,\;
 [r_2,r_3]=i\left(1+j_1^2{\bf r}(j) \right)\frac{1}{J_{III}}\tanh J_{III}v
  \Biggl\},
\end{displaymath} 
\begin{displaymath} 
 S_v^{3}(j;\sigma_{IV})= \Biggl\{
 [r_1,r_2]=[r_2,r_3]=0,\;
 [r_1,r_3]=i\left(1+j_1^2{\bf r}(j) \right)\frac{1}{J_{IV}}\tanh J_{IV}v,
 \Biggl\},
\end{displaymath} 
\begin{equation}
 S_v^{3}(j;\sigma_{V})= \Biggl\{
 [r_1,r_3]=[r_2,r_3]=0,\;
 [r_1,r_2]=i\left(1+j_1^2{\bf r}(j) \right)\frac{1}{J_{V}}\tanh J_{V}v,
 \Biggl\},
\label{4v-1}
\end{equation}
where
$
{\bf r}(j)=r_1^2 + j_2^2r_2^2 + j_2^2j_3^2r_3^2,\;
J_{III}=j_1^2j_2^2j_3,\; J_{IV}=j_1^2j_2j_3,\; J_{V}=j_1^2j_2.
$

All quantum orthogonal spheres $S_v^{3}(j;\sigma) $ can be  divided into two classes
relative to their properties under nilpotent values of contraction parameters.
These properties depend on transformation of deformation parameter and are different for minimal first order multipliers $J_0, J_{I}, J_{II}$ and for
non-minimal  multipliers $ J_{III}, J_{IV}, J_{V}.$
Let us regard these two classes separately.

For minimal  multipliers all quantum analogs of 3-dimensional space with zero curvature $(j_1=\iota_1)$ are isomorphic and can be obtained from
\begin{equation}
 S_v^{3}(\iota_1;\sigma_0)= \Biggl\{
 [r_1,r_2]=0,\;  [r_3,r_1]=ivr_1,\; [r_3,r_2]=ivr_2
 \Biggl\}
 \label{4v-2}
\end{equation}
by  permutations of generators $r_k,\; k=1,2,3.$

For $j_2=\iota_2$ the  commutative space has 1-dimensional base $\left\{r_1\right\}$
and 2-dimensional fiber $\left\{r_2, r_3\right\}$.
Corresponding quantum space
\begin{equation}
 S_v^{3}(\iota_2;\sigma_0)= \Biggl\{
 [r_1,r_2]=0,\;  [r_3,r_m]=ivr_m(1+j_1^2r_1^2),\; m=1,2
 \Biggl\}
 \label{4q-8}
\end{equation}
is transformed to $ S_v^{3}(\iota_2;\sigma_{I})$ by the substitution $2\rightarrow 3 $ and vice versa. Both spaces have noncommutative fiber. New quantum deformation with  commutative fiber is given by
\begin{equation}
 S_v^{3}(\iota_2;\sigma_{II})= \Biggl\{
 [r_2,r_3]=0,\;  [r_1,r_m]=i(r_m + j_1^2r_1r_mr_1)\frac{1}{j_1}\tanh j_1v,\; m=2,3
 \Biggl\}.
 \label{4q-9}
\end{equation}

When $j_3=\iota_3$ there are three nonisomorphic quantum spaces:
one with commutative base $\left\{r_1, r_2\right\}$
\begin{equation}
 S_v^{3}(\iota_3;\sigma_{0})= \Biggl\{
 [r_1,r_2]=0,\;  [r_3,r_m]=ivr_m\left(1 + j_1^2(r_1^2 +j_2^2r_2^2)\right),\; m=1,2
 \Biggl\}
 \label{4q-10}
\end{equation}
 and two with noncommutative base:
$S_v^{3}(\iota_3;\sigma_{I})$, which has commutation relations
(\ref{4q-25}), where
\begin{displaymath} 
r_2-\hat{r}_2=ij_1^2\hat{r}_1 r_1
 \frac{1}{J_I}\sinh J_Iv
\end{displaymath}
and
$S_v^{3}(\iota_3;\sigma_{II})$, which has commutation relations
  (\ref{4q-16}), where
 \begin{displaymath} 
r_1-\hat{r}_1=ij_1^2j_2^2\hat{r}_2 r_2
 \frac{1}{J_{II}}\sinh J_{II}v.
\end{displaymath}

In the case of nonminimal multipliers all quantum deformations of the Euclid space are isomorphic
to
\begin{equation}
 S_v^{3}(\iota_1;\sigma_{V})= \Biggl\{
 [r_1,r_3]= [r_2,r_3]=0,\; [r_1,r_2]=iv
 \Biggl\}
 \label{4q-11}
 \end{equation}
and are the simplest ones.

For $j_2=\iota_2$ two quantum spaces with commutative fiber are isomorphic
\begin{equation}
 S_v^{3}(\iota_2;\sigma_{V})= \Biggl\{
 [r_1,r_3]= [r_2,r_3]=0,\; [r_1,r_2]=iv(1+j_1^2r_1^2)
 \Biggl\}\cong S_v^{3}(\iota_2;\sigma_{IV}),
 \label{4q-12}
\end{equation}
but the quantum space with noncommutative fiber
\begin{equation}
 S_v^{3}(\iota_2;\sigma_{III})= \Biggl\{
 [r_1,r_2]= [r_1,r_3]=0,\; [r_2,r_3]=iv(1+j_1^2r_1^2)
 \Biggl\}
 \label{4q-13}
\end{equation}
presents new quantum deformation. In spite of the same commutation relations (\ref{4q-12}) and
(\ref{4q-13}) the quantum spaces $S_v^{3}(\iota_2;\sigma_{V})$ and $ S_v^{3}(\iota_2;\sigma_{III})$ need be regarded as different one because of their different fibering.

For $j_3=\iota_3$ on the contrary two quantum spaces with commutative base are isomorphic
\begin{equation}
 S_v^{3}(\iota_3;\sigma_{III})= \Biggl\{
 [r_1,r_2]= [r_2,r_3]=0,\; [r_1,r_3]=iv\left(1+j_1^2(r_1^2+j_2^2r_2^2)\right)
 \Biggl\}\cong
 S_v^{3}(\iota_3;\sigma_{IV}).
 \label{4q-14}
 \end{equation}
New quantum deformation with noncommutative base is given by
\begin{equation}
 S_v^{3}(\iota_3;\sigma_{V})= \Biggl\{
  [r_1,r_2]=i\left(1+j_1^2(r_1^2+j_2^2r_2^2)\right)  \frac{1}{j_1^2j_2}\tanh j_1^2j_2v, \;
 %
  [r_1,r_3]= [r_2,r_3]=0
 \Biggl\}.
 \label{4q-15}
 \end{equation}
Let us stress that deformation parameter remain untouched under this last contraction (\ref{4q-15}).

Physically quantum spaces (\ref{4q-8}),(\ref{4q-9}),(\ref{4q-12}),(\ref{4q-13} with $j_2=\iota_2$ can be interpreted as quantum analogs of the $(1+2)$ nonrelativistic  Newton kinematics of constant curvature (zero curvature or Galilei kinematics, when $j_1=\iota_1$).

 \section{Conclusion}

The quantum Cayley-Klein spaces of constant curvature $O_v^N(j;\sigma)$ are uniformly obtained from
the quantum Euclidean space $O_q^N$ in Cartesian coordinates
by the standard trick with real, complex, and nilpotent  numbers, using a q-analog of Beltrami coordinates.
The transformation of the quantum deformation parameter $Z=Jv$ under contraction is the important ingredient of the noncommutative quantum groups and noncommutative quantum spaces. Unlike previous papers on this subject
 contraction parameters of the second power  are included in the multiplier $J$, what make
all  contractions admissible.
The different combinations of quantum structure
and Cayley-Klein scheme of contractions and analytical continuations are described
with the help of permutations $\sigma. $
As a result  the   quantum orthogonal deformations of three and four dimensional Caley-Klein vector spaces
as well as those  
 of two and three dimensional constant curvature spaces
are obtained.

For three  dimensional Caley-Klein vector spaces
we have found two nonisomorphic quantum deformations (\ref{31}),(\ref{3q-3}) of semi-Euclidean space with one 
 dimensional base and two  dimensional fiber, as well as
two different deformations  (\ref{32}),(\ref{2'}) of semi-Euclidean space with two  dimensional base and one  dimensional fiber.
Four  dimensional Caley-Klein vector spaces have 
two  quantum deformations (\ref{bn-0}) of commutative fibered space with one dimensional base and three  dimensional fiber,
three nonisomorphic  deformations (\ref{bn-1}) of semi-Euclidean space with two dimensional base and two  dimensional fiber, as well as
two  deformations  (\ref{bn-2}) of fibered  space with three  dimensional base and one  dimensional fiber.

Concerning  spaces of constant curvature there are two nonequivalent deformations (\ref{3q-7}),(\ref{1-3q-7}) of Euclidean plane and three deformations (\ref{3q-8}),(\ref{5'}),(\ref{2-3q-7}) 
of cylinder or Newton plane.
In the dimensional three we have found two quantum deformations (\ref{4v-2}),(\ref{4q-11}) of Euclidean space,
four different deformations 
(\ref{4q-8}),(\ref{4q-9}),(\ref{4q-12}),(\ref{4q-13})
of semi-Riemannian space with one dimensional base and two  dimensional fiber and five nonisomorphic quantum  deformations (\ref{4q-25}),(\ref{4q-16}),(\ref{4q-10}),(\ref{4q-14}),(\ref{4q-15})
of semi-Riemannian space with two dimensional base and one  dimensional fiber.

This demonstrate a wide variety of the quantum deformations of of the fibered semi-Riemannian spaces.
One of their  remarkable property is that for some of them commutation relations of  generators are proportional to a {\it numbers} (\ref{1-3q-7}),(\ref{4q-11}) instead of generators, i.e. the {\it simplest} possible deformations are realized.
The unique quantum deformation of the rigid algebraic structure of   simple  Lie groups and Lie algebras \cite{FRT}
is transformed into the spectrum of nonisomorphic deformations of the more flexible contracted structure of non-semisimple Lie groups  and associated noncommutative spaces.

\section{Acknowledgments}
This work was supported by the program "Fundamental problems of nonlinear dynamics" of the Russian Academy of Sciences.

\end{document}